\documentclass[prb,fleqn,12pt,showpacs]{revtex4}
\usepackage{epsfig}
\usepackage{graphicx}
\usepackage{amsfonts}
\begin{document} 
\title{Correlation effects in a band ferromagnet: \\
spin-rotationally-symmetric study with \\
self-energy and vertex corrections}
\author{Avinash Singh}
\email{avinas@iitk.ac.in} 
\affiliation{Max-Planck-Institut f\"{u}r Physik Komplexer Systeme, N\"{o}thnitzer str. 38, D-01187 Dresden}
\affiliation{Department of Physics, Indian Institute of Technology Kanpur - 208016}
\begin{abstract}
Quantum corrections to the transverse spin-fluctuation propagator are obtained 
by including self-energy and vertex corrections to first order 
within a spin-rotationally-symmetric inverse-degeneracy $(1/{\cal N})$ expansion scheme
which preserves the Goldstone mode order by order.
A correlation-induced exchange-energy correction
is shown to yield the dominant quantum reduction in the spin stiffness,
providing a quantitative understanding of the ferromagnetic-state stability
in terms of simple lattice-dependent features of energy-band dispersion.
The quantum reduction factor $U/W$ 
highlights the subtlety in the characteristic competition in a band 
ferromagnet between interaction $U$ and bandwidth $W$. 
\end{abstract}
\maketitle
\newpage
\section{Introduction}
The continuous spin-rotation symmetry of a magnetic system is manifested, 
in the spontaneously broken-symmetry state, 
in gapless spin-wave excitations in accordance with the Goldstone theorem,\cite{goldstone} 
the zero-energy infinite-wavelength 
mode simply corresponding to a uniform rotation of all spins. 
The gapless spin-wave spectrum has particularly important consequences for
low-dimensional ($D=1,2$) magnetic systems, where transverse spin fluctuations
diverge at any finite temperature, resulting in absence of long-range magnetic
order in accordance with the Mermin-Wagner theorem,\cite{mermin}
and exponentially large spin-correlation length in two dimensions.

A quantitative determination of the spin-wave spectrum also allows for 
various ordering, dimension, and lattice-specific investigations ---
finite-temperature spin dynamics and reduction of magnetic order 
with temperature due to thermal excitation of spin waves,\cite{prl,twodim,mml}
estimation of the magnetic transition temperature $T_c$ from the 
broken-symmetry side,\cite{3daf,rumsch}     
competing interactions and magnetic stability,\cite{instab,tri}
and dispersion of magnetic and electronic excitations in solids
as inferred from inelastic neutron-scattering and angle-resolved photoemission 
studies,\cite{highres,self,selftri,holmium_neutron,holmium}
References given above illustrate recent spin-wave applications
to the nearly square- and triangular-lattice antiferromagnets
such as cuprates ($\rm La_2CuO_4$), multiferroics ($\rm HoMnO_3$), 
and organic systems $\rm \kappa-(BEDT-TTF)_2 X$,
as well as ferromagnets (Fe,Ni) and magnetic multilayers (Fe/Cr) 
exhibiting giant magneto-resistance.

For band ferromagnets such as Fe and Ni, 
there have been extensive inelastic neutron-scattering studies
in relation with calculations for transverse spin fluctuations
in the random phase approximation (RPA),
which is the lowest-order treatment in which 
spin rotation symmetry and Goldstone mode are preserved. 
Various spin-wave features such as isotropy, stiffness constant, 
damping and disappearance at higher energy 
due to interaction with the continuum of Stoner excitations,
persistence for $T>T_c$,
temperature dependence of dispersion etc. 
have been discussed,\cite{lynn1,mook,lynn2,linewidth,damping}
and also quantitatively compared with RPA calculations 
using realistic band structure.\cite{cooke1,cooke2}
However, despite the extensive study of magnetic excitations 
in metals and alloys over the years,\cite{gautier}
a spin-rotationally-symmetric extension of RPA 
including self-energy and vertex corrections 
has not been carried out quantitatively for fcc-type lattices
with respect to ferromagnetic-state stability, 
spin-wave and Stoner excitations, damping, and transition temperature.

Some of the related developments beyond RPA are summarized below.
The effect of large spin fluctuations in a nearly ferromagnetic Fermi liquid has been
studied in the context of spin waves in He$^3$ 
within an extension of the paramagnon model beyond RPA,
where ambiguities of the paramagnon model were shown to be resolved.\cite{ma}
In the context of self-energy corrections in a band ferromagnet, 
the importance of vertex corrections in restoring the spin-rotation symmetry 
and Goldstone mode has been recognized at a formal level,
and a Ward identity connecting vertex corrections 
to self-energy corrections has been derived.\cite{hertz}
While spin-wave excitations for arbitrary wave vector were not quantitatively discussed,
the spin-wave stiffness constant was shown to be reduced from its RPA value, 
and also compared with earlier studies\cite{edwards1,edwards2} 
in the context of stability of the ferromagnetic state.\cite{kanamori}
A variational approach has been used to improve the RPA result for 
the energy of long wave length spin-wave modes.\cite{roth} 
A spin-wave damping term proportional to $q^6$ 
due to scattering off particle-hole excitations
has been obtained for a parabolic band.\cite{thompson}
Recently self-energy corrections have been incorporated in a modified RPA approach, 
although the $q,\omega$-dependence of vertex corrections was not included.\cite{rumsch}

In this paper we provide a concrete extension beyond RPA 
for the transverse spin-fluctuation propagator.
We make use of the inverse-degeneracy ($1/{\cal N}$) expansion
within the generalized ${\cal N}$-orbital Hubbard model,\cite{quantum} which provides a 
systematic diagrammatic scheme for incorporating quantum corrections 
while preserving spin-rotation symmetry and hence the Goldstone mode order-by-order.
This spin-rotationally-symmetric scheme has been applied earlier to examine 
quantum corrections in the antiferromagnetic state of the Hubbard model.\cite{quantum}
The diagrams include self-energy and vertex corrections,
and physically incorporate effects such as quasiparticle damping, spectral-weight transfer
and coupling of spin and charge fluctuations.
We consider the special case of a saturated band ferromagnet,
in which the absence of minority-spin particle-hole fluctuations
results in relative simplification. 

Owing to its intrinsically strong-coupling nature,
band ferromagnetism has been recognized as a fairly challenging problem, 
particularly with respect to the estimation of Curie temperature for the Hubbard model,
although considerable progress has been achieved in the recent past.\cite{nolting_book} 
Competition between band and interaction energies,
separation of moment-melting and moment-disordering temperature scales due to 
strong correlation, and presence of charge fluctuations even in the broken-symmetry state
due to partially filled band(s) are some of the non-trivial elements involved.
Ferromagnetism in the Hubbard model on fcc and bcc lattices has been recently investigated using several different approaches, such as the dynamical mean field theory (DMFT),\cite{ulmke} by incorporating spin and charge fluctuations in the correlated paramagnet using the fluctuation-exchange (FLEX) and the two-particle self consistent (TPSC) approximations,\cite{arita} by systematically improving the self energy,\cite{nolting_book} and a modified RPA scheme.\cite{rumsch} 
Ferromagnetism in a diluted Hubbard model has also been investigated recently,\cite{dms,ukondo} which is of interest in the context 
of carrier-mediated ferromagnetism in diluted magnetic semiconductors
such as $\rm Ga_{1-x} Mn_x As$. 

Incorporating only the local (Ising) spin excitations, the DMFT approach ignores long-wavelength spin fluctuations and the ${\bf k}$-dependence of self energy.
FLEX incorporates self-energy corrections, 
but ignores vertex corrections of the same order, 
thereby breaking the spin-rotation symmetry. 
Both DMFT and FLEX are hence not in accordance with the Mermin-Wagner theorem. 
While self-energy corrections in the broken-symmetry state were incorporated in the modified RPA approach,\cite{rumsch} the momentum-energy dependence of vertex corrections 
was not considered.
While FLEX, DMFT, and RPA results for the behaviour of Curie temperature with band filling
are found to be qualitatively similar,
appreciable quantitative differences\cite{arita,nolting_book,rumsch} 
clearly highlight the need for a spin-rotationally-symmetric extension.

We consider the generalized $\cal N$-orbital Hubbard model\cite{quantum} 
\begin{equation}
H=-t\sum_{\langle ij \rangle, \sigma,\alpha}
(a_{i\sigma \alpha}^{\dagger} a_{j \sigma \alpha} + {\rm H.c.}) +
\frac{1}{\cal N} \sum_{i, \alpha, \beta}
(U_1 a^{\dagger}_{i \uparrow \alpha} a_{i \uparrow \alpha}
a^{\dagger}_{i \downarrow \beta} a_{i \downarrow \beta} 
+ 
U_2 a^{\dagger}_{i \uparrow \alpha} a_{i \uparrow \beta}
a^{\dagger}_{i \downarrow \beta} a_{i \uparrow \alpha}) \; ,
\end{equation}
where $\alpha,\beta$ refer to the degenerate orbital indices
and the factor $1/\cal N$ is included to render the energy density finite
in the ${\cal N} \rightarrow \infty$ limit. 
In the isotropic limit $U_1=U_2=U$, 
the two interaction terms 
(density-density and exchange-type with respect to orbital indices) 
are together equal to $U(-{\bf S}_i . {\bf S}_i + n_i ^2)$
in terms of the total spin ${\bf S}_i \equiv \sum_\alpha \psi_{i\alpha}^\dagger 
({\mbox{\boldmath $\sigma$}}/2) \psi_{i\alpha}$ and charge 
$n_i \equiv \sum_\alpha \psi_{i\alpha}^\dagger ({\bf 1}/2) \psi_{i\alpha}$ operators,
and the Hamiltonian is therefore explicitly spin-rotationally symmetry.

\section{Transverse spin fluctuations}
The transverse spin-fluctuation propagator in the broken-symmetry state,
which describes both collective spin-wave and particle-hole Stoner excitations,
is given by  
\begin{equation}
\chi ^{-+} ({\bf q},\omega) =
i \int dt  \; e^{i\omega (t-t')} \sum_\beta \sum_j e^{i{\bf q}.({\bf r}_i - {\bf r}_j)} 
\langle \Psi_{\rm G} |{\rm T}[
S_{i\alpha} ^- (t) S_{j\beta} ^+ (t')]\Psi_{\rm G} \rangle 
\end{equation}
in terms of the fermion spin-lowering and -raising operators 
$S^\mp = \Psi^\dagger (\sigma^\mp/2) \Psi$.
The spin-fluctuation propagator can be expressed as 
\begin{equation}
\chi^{-+}({\bf q},\omega) = \frac{\phi({\bf q},\omega)}
{1-U\phi({\bf q},\omega)}
\end{equation}
in terms of the exact irreducible propagator $\phi({\bf q},\omega)$,
which incorporates all self-energy and vertex corrections. 
The inverse-degeneracy expansion\cite{quantum} 
\begin{equation}
\phi = \phi^{(0)} + \left ( \frac{1}{\cal N} \right ) \phi^{(1)} 
+ \left (\frac{1}{\cal N} \right )^2 \phi^{(2)} + ...
\end{equation}
systematizes the diagrams in powers of the expansion parameter $1/{\cal N}$ which,
in analogy with $1/S$ for quantum spin systems, plays the role of $\hbar$.
As only the "classical" term $\phi^{(0)}$ survives in the $\cal N \rightarrow \infty$ limit, 
the RPA ladder series $\chi^0 ({\bf q},\omega)/1-U \chi^0 ({\bf q},\omega)$ 
(with interaction $U_2$) 
amounts to a classical-level description of non-interacting spin-fluctuation modes.
The bare antiparallel-spin particle-hole propagator
\begin{equation}
\phi^{(0)}({\bf q},\omega) \equiv  \chi^0 ({\bf q},\omega) =
\sum_{\bf k} \frac{1}
{\epsilon_{\bf k - q}^{\downarrow +} - \epsilon_{\bf k}^{\uparrow -} + \omega -i \eta}
\; ,
\end{equation}
where $\epsilon_{\bf k}^\sigma = \epsilon_{\bf k} - \sigma \Delta$ 
are the Hartree-Fock ferromagnetic band energies, 
$2\Delta = mU$ is the exchange band splitting,
and the superscript $+(-)$ refer to particle (hole) states above (below) 
the Fermi energy $\epsilon_{\rm F}$.
For the saturated ferromagnet, 
the magnetization $m$ is equal to the particle density $n$. 

As collective spin-wave excitations are represented by poles in (3), 
spin-rotation symmetry requires that $\phi = 1/U$ for $q,\omega=0$,
corresponding to the Goldstone mode.
Since the zeroth-order term $\phi^{(0)}$ already yields exactly $1/U$ for $q,\omega=0$, 
the sum of the remaining terms must exactly vanish in order to 
preserve the Goldstone mode. For this cancellation to hold for arbitrary $\cal N$,
each higher-order term $\phi^{(n)}$ in the expansion (4) must individually vanish,
implying that spin-rotation symmetry is preserved order-by-order,
as expected from the spin-rotationally-invariant form 
$(U/{\cal N}){\bf S}_i.{\bf S}_i$ 
of the interaction term in the generalized Hubbard model.\cite{quantum} 
We evaluate the order $1/{\cal N}$ diagrams in $\phi^{(1)}$
and explicitly show the exact cancellation for $q,\omega=0$.
 
We consider the relatively simpler case of a saturated ferromagnet in which the 
minority-spin ($\downarrow$) band is pushed above the Fermi energy due to Coulomb repulsion,
resulting in the absence of any minority-spin particle-hole processes.
In this case the effective antiparallel-spin interaction at order $1/\cal N$ 
reduces to the bare Hubbard interaction $U$ 
and the effective parallel-spin interaction 
reduces to a single term involving the majority-spin ($\uparrow$) particle-hole bubble.
Generally these effective interactions involve a series of bubble diagrams,
with even and odd number of bubbles, respectively.

\begin{figure}
\vspace*{-0mm}
\hspace*{-0mm}
\psfig{figure=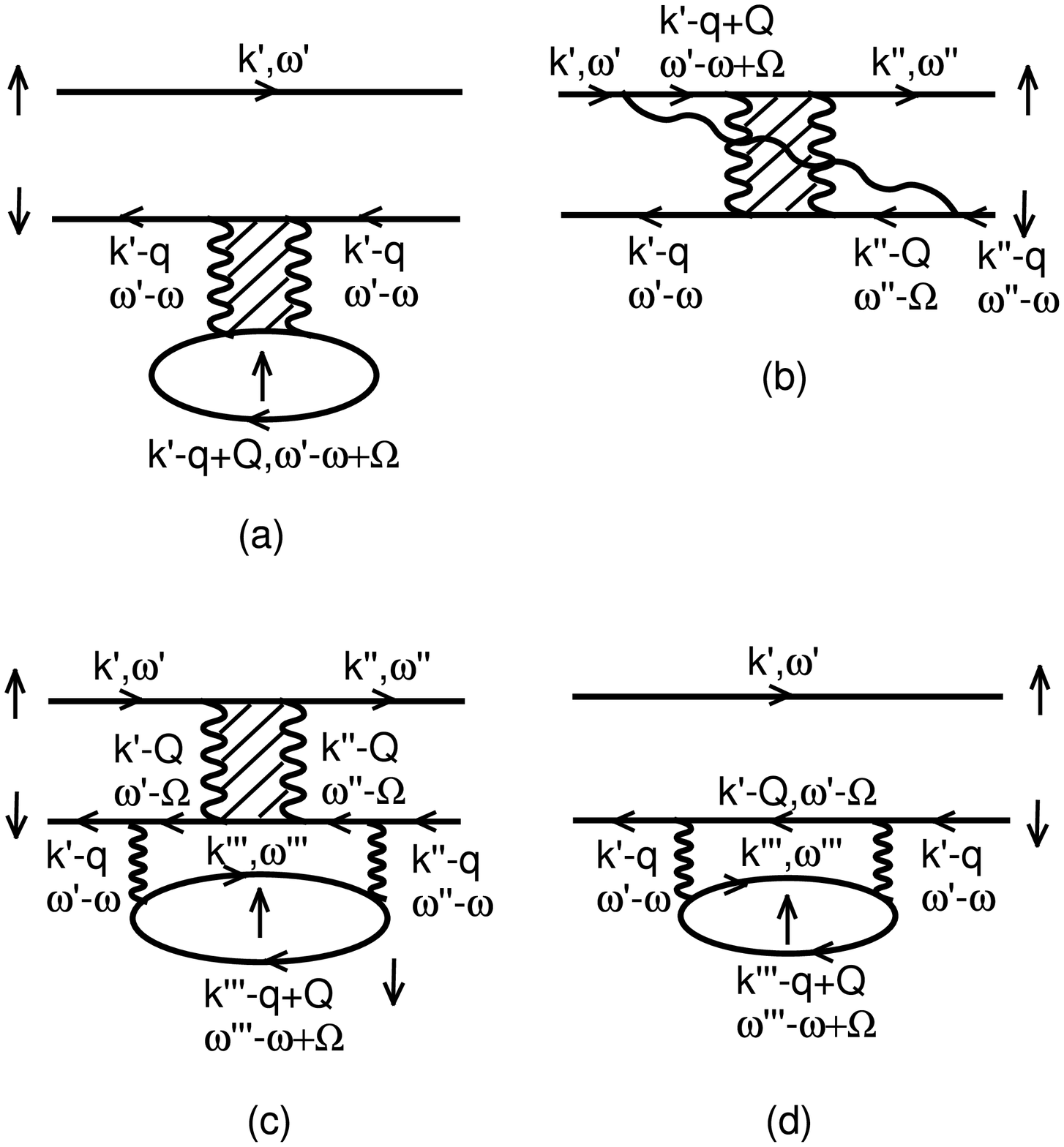,width=100mm}
\vspace{-0mm}
\caption{The first-order quantum corrections to the irreducible particle-hole 
propagator $\phi({\bf q},\omega)$.}
\end{figure}

The order $1/{\cal  N}$ diagrams for the irreducible particle-hole propagator 
$\phi({\bf q},\omega)$ are shown in Figure 1. 
The hatched part in diagram (a) represents the RPA ladder sum
$U^2 \chi^{+-}_{\rm RPA}(-{\bf Q},-\Omega)$ with interaction $U_2$, where 
\begin{equation}
\chi^{+-}_{\rm RPA}(-{\bf Q},-\Omega) = \chi^{-+}_{\rm RPA}({\bf Q},\Omega) 
= \frac{\chi^0 ({\bf Q},\Omega)}{1-U\chi^0 ({\bf Q},\Omega)}
\end{equation}
has purely advanced (retarded) character with respect to $\Omega$ ($-\Omega$),
and includes both spin-wave and Stoner excitations. 
Diagram (a) involves a self-energy correction to the $\downarrow$-spin particle 
which transfers spectral weight down from the $\downarrow$-spin band to the 
$\uparrow$-spin band (above $\epsilon_{\rm F}$), 
and yields a positive correction to $\phi$.
Diagrams (b) and (c) represent vertex corrections,
where the hatched part represents $U/1-U\chi^0 ({\bf Q},\Omega)$, 
the RPA ladder series starting with a single interaction line $U$. 
In diagram (b) the single opposite-spin {\em particle-particle} interaction $(U_1)$
reduces the $\downarrow$-spin particle --- $\uparrow$-spin hole correlation, 
yielding a negative correction to $\phi$,
whereas two such $(U_1)$ interactions in (c) and (d) yield positive corrections.

A coupling between spin and charge fluctuations is indicated by 
the $\uparrow$-spin particle-hole bubble, present explicitly in diagrams (c) and (d)
and implicitly in (a) and (b).
It is the availability of unoccupied $\uparrow$-spin states 
for partial band filling $n < 1$ which 
allows for the different processes (a)-(d),
either involving spin flip due to spin-wave coupling (a) or 
the fermion-fermion scattering due to on-site Coulomb interaction (b,c,d).
Indeed all these quantum corrections identically vanish for a completely filled 
$\uparrow$-spin band. This coupling between spin and charge fluctuations is a unique 
feature of the quantum corrections and is absent at the classical level. 

Integrating out the fermion frequency-momentum modes, 
the order $1/{\cal  N}$ quantum corrections to the irreducible particle-hole
propagator $\phi({\bf q},\omega)$ are obtained as:
\begin{eqnarray}
\phi^{(a)} ({\bf q},\omega) &=& U^2 \sum_{\bf Q} \int \frac{d\Omega}{2\pi i}
\left \{ \frac{\chi^0 ({\bf Q},\Omega)}{1-U\chi^0 ({\bf Q},\Omega)} \right \}
\nonumber \\ 
&.&
\sum_{\bf k'}
\left (
\frac{1}
{\epsilon_{\bf k' - q}^{\downarrow +} - \epsilon_{\bf k'}^{\uparrow -} + \omega -i \eta}
\right )^2 
\left (
\frac{1}
{\epsilon_{\bf k' - q + Q}^{\uparrow +} - \epsilon_{\bf k'}^{\uparrow -} 
+ \omega - \Omega - i \eta}
\right )
\end{eqnarray}

\begin{eqnarray}
\phi^{(b)} ({\bf q},\omega) &=&  -2U^2 \sum_{\bf Q} \int \frac{d\Omega}{2\pi i}
\left \{ \frac{1}{1-U\chi^0 ({\bf Q},\Omega)} \right \}
\nonumber \\ 
&.&
\sum_{\bf k'}
\left (
\frac{1}
{\epsilon_{\bf k' - q}^{\downarrow +} - \epsilon_{\bf k'}^{\uparrow -} + \omega -i \eta}
\right )
\left (
\frac{1}
{\epsilon_{\bf k' - q + Q}^{\uparrow +} - \epsilon_{\bf k'}^{\uparrow -} 
+ \omega - \Omega - i \eta}
\right )
\nonumber \\ 
&.&
\sum_{\bf k''}
\left (
\frac{1}
{\epsilon_{\bf k'' - q}^{\downarrow +} - \epsilon_{\bf k''}^{\uparrow -} + \omega -i \eta}
\right )
\left (
\frac{1}
{\epsilon_{\bf k'' - Q}^{\downarrow +} - \epsilon_{\bf k''}^{\uparrow -} + \Omega -i \eta}
\right )
\end{eqnarray}

\begin{eqnarray}
\phi^{(c)} ({\bf q},\omega) &=& U^3 \sum_{\bf Q} \int \frac{d\Omega}{2\pi i}
\left \{ \frac{1}{1-U\chi^0 ({\bf Q},\Omega)} \right \}
\nonumber \\
&.&
\sum_{\bf k'}
\left (
\frac{1}
{\epsilon_{\bf k' - q}^{\downarrow +} - \epsilon_{\bf k'}^{\uparrow -} + \omega -i \eta}
\right )
\left (
\frac{1}
{\epsilon_{\bf k' - Q}^{\downarrow +} - \epsilon_{\bf k'}^{\uparrow -} + \Omega -i \eta}
\right )
\nonumber \\ 
&.& 
\sum_{\bf k''}
\left (
\frac{1}
{\epsilon_{\bf k'' - q}^{\downarrow +} - \epsilon_{\bf k''}^{\uparrow -} + \omega -i \eta}
\right )
\left (
\frac{1}
{\epsilon_{\bf k'' - Q}^{\downarrow +} - \epsilon_{\bf k''}^{\uparrow -} + \Omega -i \eta}
\right )
\nonumber \\ 
&.&
\sum_{\bf k'''}
\left (
\frac{1}
{\epsilon_{\bf k''' - q + Q}^{\uparrow +} - \epsilon_{\bf k'''}^{\uparrow -} 
+ \omega - \Omega - i \eta}
\right )
\end{eqnarray}

\begin{eqnarray}
\phi^{(d)} ({\bf q},\omega) &=& U^2 \sum_{\bf Q} \int \frac{d\Omega}{2\pi i}
\sum_{\bf k'}
\left (
\frac{1}
{\epsilon_{\bf k' - q}^{\downarrow +} - \epsilon_{\bf k'}^{\uparrow -} + \omega -i \eta}
\right )^2
\left (
\frac{1}
{\epsilon_{\bf k' - Q}^{\downarrow +} - \epsilon_{\bf k'}^{\uparrow -} + \Omega -i \eta}
\right )
\nonumber \\ 
&.& 
\sum_{\bf k'''}
\left (
\frac{1}
{\epsilon_{\bf k''' - q + Q}^{\uparrow +} - \epsilon_{\bf k'''}^{\uparrow -} 
+ \omega - \Omega - i \eta}
\right )
\end{eqnarray}

In the infinite-wavelength limit ($q\rightarrow 0$),
the total order $1/{\cal N}$ contribution  
$\phi ^{(1)} ({\bf q},\omega)$ exactly vanishes,
as required from spin-rotation symmetry.
With $\epsilon_{\bf k}^{\downarrow +} - \epsilon_{\bf k}^{\uparrow -} = 2\Delta$,
we obtain
\begin{eqnarray}
\phi ^{(1)} ({\bf q}=0,\omega) &=& 
\phi^{(a)} + \phi^{(b)} + \phi^{(c)} + \phi^{(d)} \nonumber \\
&=& U^2 \sum_{\bf Q} \int \frac{d\Omega}{2\pi i}
\left ( \frac{1}{2\Delta +\omega} \right )^2 
\sum_{\bf k'}
\left (
\frac{1}
{\epsilon_{\bf k' + Q}^{\uparrow +} - \epsilon_{\bf k'}^{\uparrow -} 
+ \omega - \Omega - i \eta}
\right ) \nonumber \\
&.& 
\left [
\frac{\chi^0}{1-U\chi^0} - \frac{2\chi^0}{1-U\chi^0} 
+ \frac{U\chi_0 ^2}{1-U\chi^0} + \chi^0
\right ]\; ,
\end{eqnarray}
which yields identically vanishing contribution for each spin-fluctuation mode ${\bf Q}$.
We note that this mode-by-mode cancellation is quite independent of the 
spectral-weight distribution of the spin-fluctuation spectrum between collective spin-wave excitations and particle-hole Stoner excitations. 
Furthermore, the cancellation holds for all $\omega$, 
indicating no spin-wave amplitude renormalization,
as expected for the saturated ferromagnet in which there are no quantum 
corrections to magnetization. 

The coupling between spin and charge fluctuations is highlighted
by the structure of (11), where the common $\uparrow$-spin particle-hole bubble term
within the ${\bf k'}$ sum represents charge fluctuations. 
This coupling provides a spin-wave damping mechanism for a band ferromagnet,
fundamentally different from the conventional damping mechanism 
in Heisenberg insulating magnets involving decay into three spin waves.
The spin-wave decay and damping process in a band ferromagnet involves 
the imaginary part of the $\uparrow$-spin particle-hole bubble
\begin{equation}
\sum_{\bf Q} \sum_{\bf k'}
\delta(\epsilon_{\bf k' - q + Q}^{\uparrow +} - \epsilon_{\bf k'}^{\uparrow -} 
+ \omega - \Omega)
\end{equation}
which corresponds to energy conservation in the spin-wave 
(energy $-\omega =\omega_{\bf -q}$) 
decay into an intermediate-state spin wave (energy $-\Omega = \Omega_{\bf -Q}$)
plus a particle-hole excitation in the metal.
Exactly vanishing for $\omega=0$, the imaginary term increases with $|\omega|$.
The typically strongly peaked spin-wave density of states near the top end of
the spectrum yields significant spin-wave damping only for zone boundary modes.
Further investigations, including effects of disorder and diffusion pole, 
are clearly of interest in view of the observed temperature-independent linewidth 
in neutron-scattering studies.\cite{linewidth,damping}

\section{Quantum corrections to spin stiffness}
We next consider the quantum corrections for small $q$ and show that
the above exact cancellation for $q=0$ actually extends to the next order as well,
yielding spin-stiffness quantum corrections only from the higher-order surviving terms.
For analytical simplicity, 
we neglect the contribution of higher-energy Stoner excitations 
in the $\Omega$ integral in (7) - (9), 
and only consider contributions from the spin-wave pole 
\begin{equation}
\frac{\chi^0({\bf Q},\Omega)}{1-U\chi^0({\bf Q},\Omega)} 
= \frac{m_{\bf Q}}{\Omega+\Omega_{\bf Q} -i\eta} \; ,
\end{equation} 
where $m_{\bf Q}$ is the spin-wave amplitude for momentum ${\bf Q}$. 
Writing the antiparallel-spin particle-hole energy denominators as
\begin{equation}
\epsilon_{\bf k - q }^{\downarrow +} - \epsilon_{\bf k}^{\uparrow -} 
= 2\Delta[1 + (\epsilon_{\bf k - q } - \epsilon_{\bf k})/2\Delta]
\end{equation}
and expanding in powers of the small band-energy difference 
\begin{equation}
\delta \equiv -(\epsilon_{\bf k - q } - \epsilon_{\bf k})
= {\bf q}.{\mbox{\boldmath $\nabla$}} \epsilon_{\bf k} - 
\frac{1}{2}({\bf q}.{\mbox{\boldmath $\nabla$}})^2 \epsilon_{\bf k}
\end{equation}
for small $q$, 
we find that besides the zeroth-order cancellation for $q=0$, 
the first-order terms in $\delta$ also exactly cancel,
implying no quantum correction to the classical term  
$\langle {\mbox{\boldmath $\nabla$}}^2 \epsilon_{\bf k} \rangle $
in the spin-wave stiffness constant,
in accordance with the exact structure.\cite{edwards1,edwards2}  
The surviving second-order terms in $\delta$ can be written, up to order $q^2$, as
\begin{eqnarray}
\phi^{(1)}({\bf q}) &=& \frac{U^2}{(2\Delta)^4} 
\sum_{Q} m_{\bf Q} \sum_{\bf k'} 
\frac{( {\bf q}.{\mbox{\boldmath $\nabla$}} \epsilon_{\bf k'} )^2 }
{\epsilon_{\bf k' + Q}^{\uparrow +} - \epsilon_{\bf k'}^{\uparrow -} 
+ \Omega_{\bf Q}} \nonumber \\
&-& 
\frac{2U^3}{(2\Delta)^4} 
\sum_{Q} m_{\bf Q} 
\sum_{\bf k'} 
\frac{{\bf q}.{\mbox{\boldmath $\nabla$}} \epsilon_{\bf k'}}
{\epsilon_{\bf k' + Q}^{\uparrow +} - \epsilon_{\bf k'}^{\uparrow -} 
+ \Omega_{\bf Q}} 
\sum_{\bf k''} 
\frac{{\bf q}.{\mbox{\boldmath $\nabla$}} \epsilon_{\bf k''}}
{\epsilon_{\bf k'' - Q}^{\downarrow +} - \epsilon_{\bf k''}^{\uparrow -} 
- \Omega_{\bf Q} } 
\nonumber \\
&+& 
\frac{U^4}{(2\Delta)^4} 
\sum_{Q} m_{\bf Q} 
\left ( 
\sum_{\bf k'} \frac{{\bf q}.{\mbox{\boldmath $\nabla$}} \epsilon_{\bf k'}}
{\epsilon_{\bf k' - Q}^{\downarrow +} - \epsilon_{\bf k'}^{\uparrow -} 
- \Omega_{\bf Q} } \right )^2
\sum_{\bf k'''} 
\frac{1}
{\epsilon_{\bf k''' + Q}^{\uparrow +} - \epsilon_{\bf k'''}^{\uparrow -} 
+ \Omega_{\bf Q} } \; ,
\end{eqnarray}
where we have set $q,\omega =0$ in the energy denominators 
as all three terms are already explicitly second order in $q$.
Incorporating the energy cost of spin twisting,
the three terms in (16) represent exchange-type processes
involving $\uparrow$-spin particle-hole (charge) excitations 
accompanied with zero, one, and two $\downarrow$-spin 
particle-particle scatterings, respectively.  
Cross terms such as $q_x q_y$ etc. in (16) identically vanish from symmetry,
leaving an isotropic momentum dependence on $q^2 = q_x ^2 + q_y ^2 + q_z ^2$. 
Equation (16) is a new result 
and incorporates all first-order ($1/{\cal N}$) quantum corrections 
to the spin-wave stiffness constant due to collective spin-wave excitations. 
Including the Stoner contribution in (7-10) again yields an exact cancellation 
of the ${\mbox{\boldmath $\nabla$}}^2 \epsilon_{\bf k}$-type terms, 
leaving only an additional $({\mbox{\boldmath $\nabla$}} \epsilon_{\bf k})^2 $-type term 
qualitatively similar to the first term in (16).\cite{spandey} 
 
Now, as ${\mbox{\boldmath $\nabla$}} \epsilon_{\bf k}$ is odd in momentum ${\bf k}$,
the second and third terms in (16) involve a partial cancellation resulting from 
the momentum summations.
Therefore, keeping the contribution of the dominant first term only, 
with an order-of-magnitude estimate for the particle-hole energy denominator as $W$,
the fermion bandwidth, the renormalized spin-wave energy in $D$ dimensions is obtained as
\begin{equation}
\omega_{\bf q} \approx  \frac{1}{D} \left [
\frac{1}{2} \langle {\mbox{\boldmath $\nabla$}}^2 \epsilon_{\bf k} \rangle 
- \frac{\langle ({\mbox{\boldmath $\nabla$}} \epsilon_{\bf k})^2 \rangle}{2\Delta}
- \frac{\langle ({\mbox{\boldmath $\nabla$}} \epsilon_{\bf k})^2 \rangle}{2\Delta}
\frac{U}{W} (1-n) \alpha \right ] q^2 \; ,
\end{equation}
where $2\Delta = nU$, 
the angular bracket $\langle \; \rangle$ represents momentum summation
normalized over the number of occupied states, 
the explicit hole density factor $(1-n)$ highlights the particle-hole process involved
which vanishes for a filled band, $\alpha$ is a band-dependent factor of order 1,
and the prefactor $1/D$ follows from hypercubic symmetry.
A straightforward calculation of the two competing terms in spin stiffness 
$\langle {\mbox{\boldmath $\nabla$}}^2 \epsilon_{\bf k} \rangle$ and  
$\langle ({\mbox{\boldmath $\nabla$}} \epsilon_{\bf k})^2 \rangle/2\Delta$
can provide a quantitative estimate of the stability of the ferromagnetic state for different lattices.

The origin and physical interpretation of the three terms in (16) are given below.
The first two terms represent classical (RPA) spin-stiffness contributions 
arising from the two fermion band curvature terms (15) substituted in (5). 
With respect to stiffness against spin twisting,
the first (positive) term represents delocalization-energy loss
which vanishes for a filled band,
whereas the second (negative) term of order $t^2/U$ represents exchange-energy gain.
The third term represents an additional exchange process involving minority-spin intermediate states
which are transferred to lower energies,
corresponding to the finite probability $1-n$ of a site being unoccupied by a majority-spin electrons.
The third term thus represents a correlation-induced quantum correction to spin stiffness.

If the normalized averages $\langle \rangle$ in (17) are essentially $n$-independent
(as for the fcc lattice with $t'=0.25$ in the low-density limit),
the band-filling dependence $(1-n)/n$ of the quantum term destabilizes the ferromagnetic 
state for small $n$, whereas the competition between the two classical terms destabilizes 
the ferromagnetic state as $n$ approaches 1,
yielding an optimization of the spin stiffness at some intermediate $n$.
Also, the quantum correction factor $U/W$ again highlights the characteristic 
competition between interaction $U$ and bandwidth $W$,
although favouring a stability condition quite opposite to the Stoner criterion.
In the low-density limit ($n \ll 1$), 
the second and third terms in (17) can be combined as 
$\langle ({\mbox{\boldmath $\nabla$}} \epsilon_{\bf k})^2 \rangle/nU_{\rm eff}$,
where the effective interaction $U_{\rm eff} = U/(1+U/W)$
approaches the bandwidth $W$ in the strong coupling limit, 
in agreement with the low-density result of Kanamori.\cite{kanamori}

\section{Conclusions}
We have investigated correlation effects on the spin dynamics in a band ferromagnet,
providing a physically transparent framework for quantitative understanding of
ferromagnetic stability in terms of simple lattice-dependent features of energy-band dispersion. 
The correlation effect arises from the minority-spin spectral-weight transfer to lower energies
corresponding to finite site-vacancy probability $(1-n)$ of majority-spin electrons,
the availability of these low-lying intermediate states 
resulting in an additional exchange-energy gain of order $t^2/U$, 
yielding a lattice-dependent destabilization of the ferromagnetic state.

The inverse-degeneracy ($1/{\cal N}$) expansion provides a fully spin-rotationally-symmetric 
scheme for incorporating self-energy and vertex corrections 
in the transverse spin-fluctuation propagator. 
First-order contributions to the irreducible particle-hole propagator were obtained,
with full momentum-energy dependence in the vertex corrections, 
and shown to have appropriate cancellations for $q=0$, small $q$, as well as finite $\omega$. 
Lattice-specific evaluations thus allow for quantitative study of 
magnetic excitations for arbitrary wave vector,
which should be especially suitable for low-dimensional systems
as the scheme is in accordance with the Mermin-Wagner theorem. 

The factor $U/W$ in the spin-stiffness quantum reduction 
highlights the subtlety in the characteristic competition in a band ferromagnet 
between interaction $U$ and bandwidth $W$. 
While Stoner criterion favours a high density of states at Fermi energy,
the quantum correction favours a large bandwidth for the stability of the 
ferromagnetic state.
Indeed, with a large density of states at one end and a broad band tail, 
the fcc-lattice band does provide optimum conditions for a stable 
ferromagnetic state within the single-band Hubbard model.
The quantum corrections also involve a coupling between spin and charge fluctuations,
resulting in an intrinsic spin-wave damping mechanism 
which fundamentally distinguishes between band and insulating ferromagnets. 
Lattice-specific calculations including the contribution of both spin-wave and Stoner excitations
in the intermediate-state spin-fluctuation spectrum are currently in progress,
and quantitative results for spin stiffness, spin-wave energy, Curie temperature,
electronic spectral function etc. will be presented elsewhere.\cite{spandey} 

\ \\
\ \\

\end{document}